\documentclass[12pt]{article}
\usepackage{epsfig,amsfonts,amssymb,amsbsy,amsbsy,array}
\def\be{\begin{equation}}\def\ee{\end{equation}}
\def\ba{\begin{eqnarray}}\def\ea{\end{eqnarray}}
\def\cvp{\raise 2pt\hbox{,}}

\def\tr{\mathop{\rm tr}\nolimits}

\def\d{{\rm d}}\def\nn{{\cal N}}
\def\suN{{\rm SU}(N)}
\def\uN{{\rm U}(N)}

\def\wt{W_{\rm tree}}

\def\La{\Lambda}
\def\la{\lambda}\def\lck{\lambda_{{\rm c},k}}

\def\mq{{\cal M}_{\rm q}}
\def\u{{\rm U}(1)}\def\uu{{\rm U}(2)}
\def\uuu{{\rm U}(3)}\def\uuuu{{\rm U}(4)}

\def\plb#1#2#3{{\it Phys.\ Lett.\ }{\bf B #1} (#2) #3}
\def\npb#1#2#3{{\it Nucl.\ Phys.\ }{\bf B #1} (#2) #3}

\def\prl#1#2#3{{\it Phys.\ Rev.\ Lett.\ }{\bf #1} (#2) #3}
\def\jhep#1#2#3{{\it J. High Energy Phys.\ }{\bf #1} (#2) #3}
\def\prd#1#2#3{{\it Phys.\ Rev.\ }{\bf D #1} (#2) #3}

\begin{document}
%
%
\pagestyle{empty}
{\parskip 0in

\hfill NEIP-03-001

\hfill LPTENS-03/01

\hfill hep-th/0301157}

\vfill
\begin{center}
{\LARGE Quantum parameter space in super Yang-Mills, II}



\vspace{0.4in}

Frank F{\scshape errari}{\renewcommand{\thefootnote}{$\!\!\dagger$}
\footnote{On leave of absence from Centre 
National de la Recherche Scientifique, Laboratoire de Physique 
Th\'eorique de l'\'Ecole Normale Sup\'erieure, Paris, France.}}
\\
\medskip
{\it Institut de Physique, Universit\'e de Neuch\^atel\\
rue A.-L.~Br\'eguet 1, CH-2000 Neuch\^atel, Switzerland}\\
\smallskip
{\tt frank.ferrari@unine.ch}
\end{center}
\vfill\noindent

In [1] (hep-th/0211069),
the author has discussed the quantum parameter space
of the ${\cal N}=1$ super
Yang-Mills theory with one adjoint Higgs
field $\Phi$, tree-level superpotential $\wt = m\Phi^{2}/2 +
g\Phi^{3}/3$, and gauge group $\uN$. In particular, full details were 
worked out for $\uu$ and $\uuu$. By discussing higher rank gauge 
groups like $\uuuu$, for which the classical parameter space has a 
large number of disconnected components,
we show that the phenomena discussed in [1] are generic. It turns out 
that the quantum space is connected. 
The classical components are related in the quantum theory either
through standard singularities with massless monopoles or by branch
cuts without going through any singularity. The branching points
associated with the branch cuts correspond to new strong
coupling singularities, which are not associated with vanishing cycles in
the geometry, and at which glueballs can become massless. 
The transitions discussed recently by 
Cachazo, Seiberg and Witten are special instances of those phenomena. 

\vfill

\medskip
%
\begin{flushleft}
January 2003
\end{flushleft}
\newpage\pagestyle{plain}
\baselineskip 16pt
\setcounter{footnote}{0}

%
\section{Introduction and review of $\uu$ and $\uuu$}

In a recent paper \cite{f1}, the powerful technology in the
calculation of exact quantum effective superpotentials
\cite{CIV,CV,DV,fer} was used for the first time to derive new physics
in $\nn=1$ supersymmetric $\uN$ gauge theories. The basic object
considered in \cite{f1} is the quantum space of parameters $\mq$. This
space is reminiscent of the quantum moduli space of theories with a
larger number of supersymmetries. The most fundamental difference is
that no massless scalar is associated with the motion on $\mq$. The
problem of calculating $\mq$ is thus very general and also occurs in
non-supersymmetric theories, as exemplified in \cite{F1}. The
space $\mq$ describes the phase diagram as well as some non-trivial
phenomena that occur in given phases.

As in \cite{f1}, the example that we consider is the $\nn=1$ theory
with gauge group $\uN$ and one adjoint Higgs field $\Phi$ with
tree-level superpotential
\begin{equation}
\label{Wtree}
\wt(\Phi) = {m\over 2}\, \Phi^{2} + {g\over 3}\, \Phi^{3}\, .
\end{equation}
Quantum mechanically, the theory depends on a single dimensionless 
parameter
\be\la = {8g^{2}\La^{2}\over m^{2}}\,\cvp\ee
where $\La$ is the dynamically generated scale that governs the UV
running of the gauge coupling constant. Weak coupling corresponds to
$\la\rightarrow 0$. In the small $\la$ region, the parameter space is
simply the union of several disconnected components, or sheets,
associated with
the various classical vacua. For example, the $\uu$ theory has five
disconnected components by taking into account chiral symmetry
breaking in the low energy gauge group. Two vacua correspond to a
classically unbroken gauge group $\uu$
with $\langle\phi\rangle_{\rm cl}=0$,
two others to similar vacua with $\langle\phi\rangle_{\rm cl}=-m/g$,
and a last vacuum corresponds to a classically unbroken gauge group
$\u\times\u$. Similarly, the $\uuu$ theory has ten weakly coupled
components, and the $\uuuu$ theory eighteen. For $\uN$,
there are $N(N^{2}+11)/6$ components
that can be labeled as
$|k_{1},k_{2};N_{1},N_{2}\rangle$. The integers $N_{1}$ and $N_{2}$,
$N_{1}+N_{2}=N$, give the number of eigenvalues of
$\langle\phi\rangle_{\rm cl}$ at zero and $-m/g$ respectively,
corresponding to a
classical breaking of the gauge group from $\uN$ down to
${\rm U}(N_{1})\times {\rm U}(N_{2})$. The integers $k_{j}$,
defined modulo $N_{j}$ and usually chosen such that
$0\leq k_{j}\leq N_{j}-1$, label the various chirally
asymmetric vacua of the low energy theory. Note that if the model were
purely classical, any two sheets of the 
parameter space would meet only at $\la =\infty$ or equivalently for
a critical $m=0$ tree-level superpotential (\ref{Wtree}).

\begin{figure}
\centerline{\epsfig{file=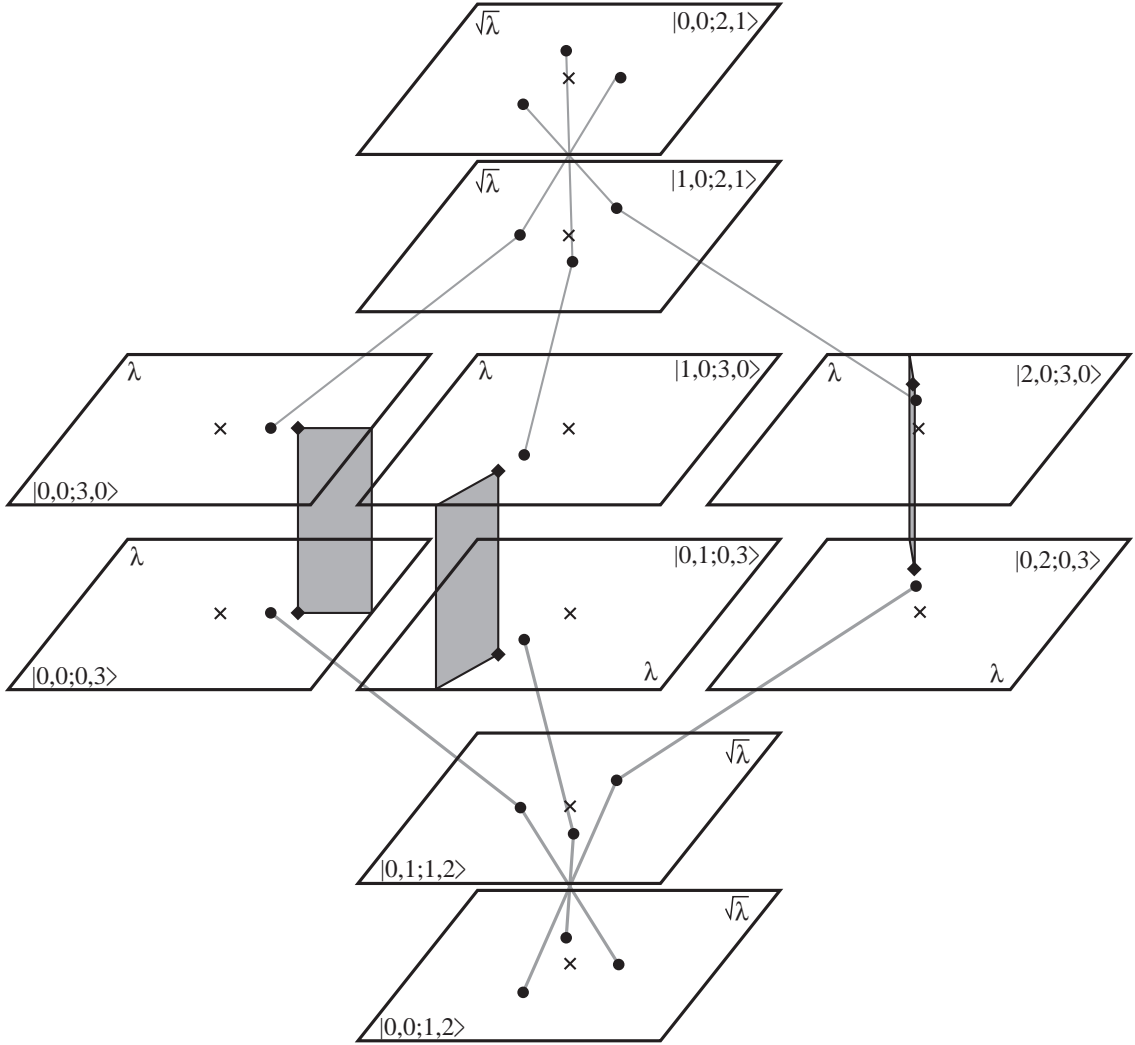,width=15cm}
}
\caption{Sketch of the quantum parameter space $\mq$ for gauge group $\uuu$. The
sheets are labeled by a state $|k_{1},k_{2};N_{1},N_{2}\rangle$ and
parametrized by $\la$ (if $N_{1}N_{2}=0$) or by
$\sqrt{\la}$ (if $N_{1}N_{2}\not =0$). The cross
denotes the classical $\la =0$ points. The black dots represent
singularities with a massless monopole that are found for
$\sqrt{\la} = 2\sqrt{2} e^{-i\pi q/3}/3$, $q$ integer. 
Each singularity of that type link three sheets, and
the gray lines represent the corresponding three-fold indentification.
The black squares at $\la = e^{-2i\pi k/3}$ represent
singularities with a massless glueball. They are the three branching
points of the three cuts joining the sheets $|k,0;3,0\rangle$ and
$|0,k;0,3\rangle$.
\label{fig}}
\end{figure}

The full quantum parameter space for $\uu$ was drawn in the figure 3
of \cite{f1}. All the relevant calculations for $\uuu$ were also
included in \cite{f1}, and we have depicted the resulting $\mq$ in
figure 1. A most notable feature is that the quantum parameter space
is connected, and this is probably the case for any gauge group $\uN$.
This connectedness results from two basic mechanisms described in
\cite{f1}.

The first mechanism corresponds to a phase transition with
massless monopole(s) and vanishing string tension(s). For example, a 
Coulomb phase and a confining phase are related in that way in figure 
1 (gray lines).
Such transitions
come with vanishing cycles in the geometry, yielding a singular matrix
model curve in the approach of \cite{DV}, or equivalently a
singular Calabi-Yau in the string theory approach of \cite{CIV}. Those 
singularities are reminiscent of the singularities on the moduli space 
of $\nn=2$ theories. 

The second mechanism involves the presence of branch cuts in parameter
space whose origin can be easily described. The expectation values of
gauge invariant chiral operators, like the $u_{k}=\tr\phi^{k}/N$, or
the glueball superfields $S_{i}$ that are the basic variables in the
string ot matrix model approaches \cite{CIV,DV}, are generically
analytic functions of the parameters. A particularly useful example
for us is
\be\label{defu} u =-{g\over m}\,\langle \tr\phi\rangle\, ,\ee
whose classical value in a vacuum $|k_{1},k_{2};N_{1},N_{2}\rangle$
is simply
\be u_{\rm cl}=u(\la =0) =N_{2}\, .\ee
The vevs are found by extremizing suitable effective quantum
superpotentials, and this amounts to solving an algebraic equation.
One then gets infinite fractional instanton series that have a finite
radius of convergence. The corresponding analytic functions have
branch cuts. For example, in the $\uN$ theory, one has \cite{f1}
\be\label{u13}u(\la)= {N\over 2}
\Bigl( 1-\sqrt{\displaystyle 1-\la e^{2i\pi k/N}}\Bigr)\ee
for the vacua $|N,0;k,0\rangle$. Equation (\ref{u13})
yields the correct classical limit $u(\la=0)=0$. By going through the
branch cut of the square root, one joins a sheet corresponding to a
different classical limit $u_{\rm cl}=N$, suitable for the
vacua $|0,N;0,k\rangle$. This demonstrates the existence of the branch
cuts in figure 1. Since going through a branch cut is a smooth
operation, components of the parameter space connected in this way
must be in the same phase. This is obviously the case for the vacua
$|N,0;k,0\rangle$ and $|0,N;0,k\rangle$ that are both confining. More
precisely, they are in the same oblique confining phase characterized
by the integer $k$ that represents the electric charge of the
condensed dyons. At the branching points, analyticity in the vev of
the chiral operator is lost. This chiral operator then overlaps with
massless degrees of freedom. This is a new kind of singularity that is
not associated with a singular geometry. In the example described by
figure 1, both $\tr\phi$ and the glueball $S$ are massless at the
branching points \cite{f1}. A last subtlety is that
different chiral operators vevs can have different analytic
structures. For example, even if two sheets are related by a branch cut 
on $\mq$, the expectation value of a given chiral operator that turns 
out to have the same
classical limit on the two sheets does not need to have any branch cut.

An interesting problem is to find general criteria for the presence of 
monopole singularities and/or branch cuts. A monopole that is not 
condensed in a vacuum $|k_{1},k_{2};N_{1},N_{2}\rangle$ can become 
massless only if
\be\label{c1} k_{1}-k_{2} \equiv 0\ {\rm mod}\ N_{1}\wedge N_{2}\, ,\ee
where $N_{1}\wedge N_{2}$ is the greatest common divisor of $N_{1}$
and $N_{2}$. This condition was derived in \cite{f1} by looking at
constraints on the possible singularities of the matrix model curve.
On the other hand, the presence of a branch cut relating two sheets is
possible only if they are in the same phase. One can use the general
analysis by 't~Hooft \cite{phooft}, based on the Wilson and 't~Hooft
loop operators, to give a criterion for
$|k_{1},k_{2};N_{1},N_{2}\rangle$ and
$|k_{1}',k_{2}';N_{1}',N_{2}'\rangle$ to be in the same phase. As
explained recently in \cite{CSW}, it is useful to introduce the
confinement index $t$. It takes values in the interval $[1,N]$ and
is defined to be the smallest integer such that the $t^{\rm th}$
tensor product of the fundamental representation of $\suN$ does not
confine. Two components can then be in the same phase only if they have the
same confinement index. Within 't~Hooft's classification scheme, one
must have \cite{CSW}
\be\label{cifor} t = N_{1}\wedge N_{2}\wedge (k_{1}-k_{2})\, ,\ee
where, by adding a multiple of $N$ if need be, $1\leq k_{1}-k_{2}\leq N$.
An interesting remark \cite{CSW} is that components in the same
phase, and in particular with the same value of $t$, can correspond to
different classically unbroken gauge groups. This is of course
possible because the notion of a broken gauge group only makes sense 
classically. Unfortunately, it appears that components with the
same values of $t$ are not necessarily connected, and more general
criteria are needed \cite{CSW}.

In the present note, we derive the quantum parameter space for the
gauge group $\uuuu$. Our motivation was to check on a rather complex
example that the phenomena described in \cite{f1} are generic, and in
particular that the results of \cite{CSW} can be understood in terms of the 
ideas reviewed above. We will
demonstrate that the eighteen components of the $\uuuu$
parameter space are all related to each other through one of the two
mechanisms described above. The final result is summarized
in figure 2. We also briefly discuss the effective
description of the parameter space with the help of the glueball
superpotentials of \cite{DV}. Finally, in the concluding section, we list a
series of open problems in the field, including remarks
on large $N$ and the Dijkgraaf-Vafa matrix model description.

\section{The case of $\uuuu$}
\setcounter{equation}{0}
\subsection{Calculations}

The calculations relevant to the ten sheets $|k,0;4,0\rangle$, 
$|0,k;0,4\rangle$ and $|k,k;2,2\rangle$ were already performed in 
\cite{f1}. Formula (\ref{u13})
shows that we have branching points at
\be\label{cl}\la=\lck = e^{-i\pi k/2}\, ,\quad 0\leq k\leq 3\, ,\ee
and associated branch cuts joining the sheets $|k,0;4,0\rangle$ and
$|0,k;0,4\rangle$. Both $\tr\phi$ and the glueball superfield $S$ are
massless at the branching points and give equally valid description of
the low energy physics. It turns out that there is also a massless
monopole at the branching point, as is actually the case for all gauge
groups ${\rm U}(2N)$ in the sheets with $N_{1}N_{2}=0$ \cite{f1}.
Through the monopole singularity, we can go to another phase
$|k,k;2,2\rangle$, reducing the confinement index from 4 to 2. More
precisely, we have a massless monopole at $\la = \pm 1$ on the sheet
$|0,0;2,2\rangle$ and a massless monopole at $\la = \pm i$ on the
sheet $|1,1;2,2\rangle$. At this stage of the analysis, the ten
disconnected sheets we started from thus appear to be grouped together
in two five-sheeted connected branches, each very similar to the
${\rm U}(2)$ quantum parameter space depicted in the figure 3 of
\cite{f1}.

The remaining eight sheets correspond to the state $|0,1;2,2\rangle$,
$|1,0;2,2\rangle$, $|k,0;3,1\rangle$ and $|0,k;1,3\rangle$. All those
sheets have confinement index 1 and are thus in a Coulomb phase
\cite{CSW}. They can a priori be related to each other through branch
cuts. Our general discussion suggests that this possibility is
realized through the existence of branch cuts in the analytic function
$u(\la)$ defined in (\ref{defu}). By taking into account the
multiplicities due to chiral symmetry breaking in the low energy gauge
groups, we expect that $u$ will satisfy a degree eight polynomial
equation $Q_{8}(u)=0$ with a classical polynomial
\be\label{Qcl} Q_{8,\,\rm cl}(u) = (u-1)^{3}(u-2)^{2}(u-3)^{3}\, .\ee
The exact quantum polynomial can be straightforwardly obtained by using 
the results of \cite{CIV,CV}. The vacua with $N_{1}N_{2}\not =0$ are 
described by the following equation,
\be\label{geq} g^{2}P_{+}(x)P_{-}(x) = (x-h_{1})^{2}(x-h_{2})^{2}
\left( x^{2}(m+g x)^{2} - 4 S g x - r\right)\, ,\ee
where
\be\label{Ppmdef} P_{\pm}(x) = \prod_{i=1}^{4}(x-x_{i}) \mp 
2\La^{4}\, ,\ee
$S$ is the glueball superfield and the $x_{i}$ are such that
\be \tr\phi^{q} = \sum_{i=1}^{4} x_{i}^{q}\, ,\quad 1\leq q\leq 4\, .\ee
The formula (\ref{geq}) yields eight equations for the eight unknown
variables $x_{i}$, $S$, $h_{1}$, $h_{2}$ and $r$. The solutions to
(\ref{geq}) corresponding to the cases where $P_{+}$, or $P_{-}$, has
two double roots yield the sheets $|k,k;2,2\rangle$. We are thus
interested in the other possibility, where $P_{+}$ and $P_{-}$ each
have one double root,
\be\label{eq1} P_{+}(x) = (x-h_{1})^{2}(x-a_{1})(x-a_{2})\, ,\quad
P_{-}(x) = (x-h_{2})^{2}(x-b_{1})(x-b_{2})\, .\ee
The matrix model curve, on which the constraint (\ref{c1}) applies, is
\be\label{MMc}y^{2}_{\rm MM}=g^{2}(x-a_{1})(x-a_{2})(x-b_{1})(x-b_{2})
= x^{2}(m+g x)^{2} - 4 S g x - r\, .\ee
By a straightforward, but tedious, algebraic elimination of all the
variables but $u =-(g/m) (x_{1}+x_{2}+x_{3}+x_{4})$, or much more
efficiently by plugging the equations in Mathematica, we get
\be\label{qeq} Q_{8}(u) = (u-1)^{3}(u-2)^{2}(u-3)^{3} + {\la^{4}\over 
64} = 0\, .\ee
The eight roots of (\ref{qeq}), for which explicit formulas 
generalizing (\ref{u13}) can be 
given, describe the eight sheets that we 
consider. It is straightforward to find the small $\la$ expansions in 
the various sheets. The good expansion parameter is $\la^{1/3}$.
For example, in a given set of conventions for 
the integers $k_{i}$, we have
\ba
&& |0,1;2,2\rangle :\quad u = 2 + \la^{2}/8 + \cdots\\
&&|0,0;3,1\rangle :\quad u = 1 + \la^{4/3}/8 + 7\la^{8/3}/384 + 
5\la^{4}/1024 + \cdots\\
&&|0,0;1,3\rangle :\quad u = 3 - \la^{4/3}/8- 7\la^{8/3}/384 - 
5\la^{4}/1024 + \cdots
\ea
The expansions for the other sheets are obtained by using $2\pi$ 
shifts of the bare $\theta$ angle, which amounts to performing the 
following changes,
\be\label{shift} |k_{1},k_{2};N_{1},N_{2}\rangle\rightarrow 
|k_{1}+1,k_{2}+1;N_{1},N_{2}\rangle\, ,\quad \la^{1/3}\rightarrow 
e^{-i\pi/6}\la^{1/3}\, .\ee
The fractional instanton series for $u$ converge 
only for $|\la|^{4}<27/4$. There are branching points for the critical 
values
\be\label{cl2}\la^{1/3}=\tilde\lck^{1/3}
= {3^{1/4}\over 2^{1/6}}\, e^{-i\pi 
k/6},\quad 0\leq k\leq 11\, ,\ee
at which two pairs of roots of the polynomial $Q_{8}$ in (\ref{qeq}) 
coincide. Those pairs correspond to the pairs of sheets 
$(|0,0;3,1\rangle,|1,0;2,2\rangle)$ and 
$(|0,0;1,3\rangle,|0,1;2,2\rangle)$ for $\la = \tilde\la_{\rm c,0}$, 
and the other cases are deduced from (\ref{shift}). The full analytic 
structure is depicted on figure 2, and shows that the eight 
classically disconnected vacua are fully connected in the quantum 
theory.

The critical points (\ref{cl2}) are the exact analogues of 
the critical points (\ref{cl}). In particular, it is possible to show 
that the glueball field $S$ is massless. This must be so because the 
small $\la$ values of $\langle S\rangle$ are different on the 
different sheets. An explicit check can be performed by computing the 
algebraic equation satisfied by $\sigma = S/(m\La^{2})$. By 
eliminating all the variables but $S$ from (\ref{geq}) we find a 
degree eight polynomial equation
\be\label{seq} R_{8}(\sigma) = \sigma^{8} - {\la^{2}\over 
64}\,\sigma^{2} + {\la^{4}\over 1024} = 0\, .\ee
The eight solutions correspond to the eight sheets, and the
singularities with massless glueballs correspond to critical values of
$\la$ for which two roots of $R_{8}$ coincide. As
expected, this occurs precisely for the values (\ref{cl2}).

The last step in the calculation of $\mq$ consists of studying 
possible phase transitions with massless monopoles relating the 
eight Coulomb sheets discussed above to the confining sheets. By 
plugging $a_{1}=a_{2}$ or $b_{1}=b_{2}$ in equations (\ref{eq1}) 
and (\ref{MMc}), we get four values of $\la$ with a massless monopole,
\be\label{monsg} \la = \la_{\rm monopole\, ,k} = (4/5)e^{-i\pi 
k/2}\, .\ee
To identify precisely on which sheets the singularities appear, one
can calculate $u$ in each cases.
For $\la = -4i /5$, we obtain $u = 2-2/\sqrt{5}$ or $u =
2+2/\sqrt{5}$. The first value is common to the sheets
$|1,0;4,0\rangle$ and $|0,0;3,1\rangle$, and the second value is
common to $|0,1;0,4\rangle$ and $|0,0;1,3\rangle$. All the other cases
are deduced from (\ref{shift}). In particular, there is no massless 
monopole point on the sheets $|1,0;2,2\rangle$ and $|0,1;2,2\rangle$, 
a fact that also follows from (\ref{c1}).

The final outcome is a fully connnected space $\mq$ depicted on figure 2.

\begin{figure}
\centerline{\epsfig{file=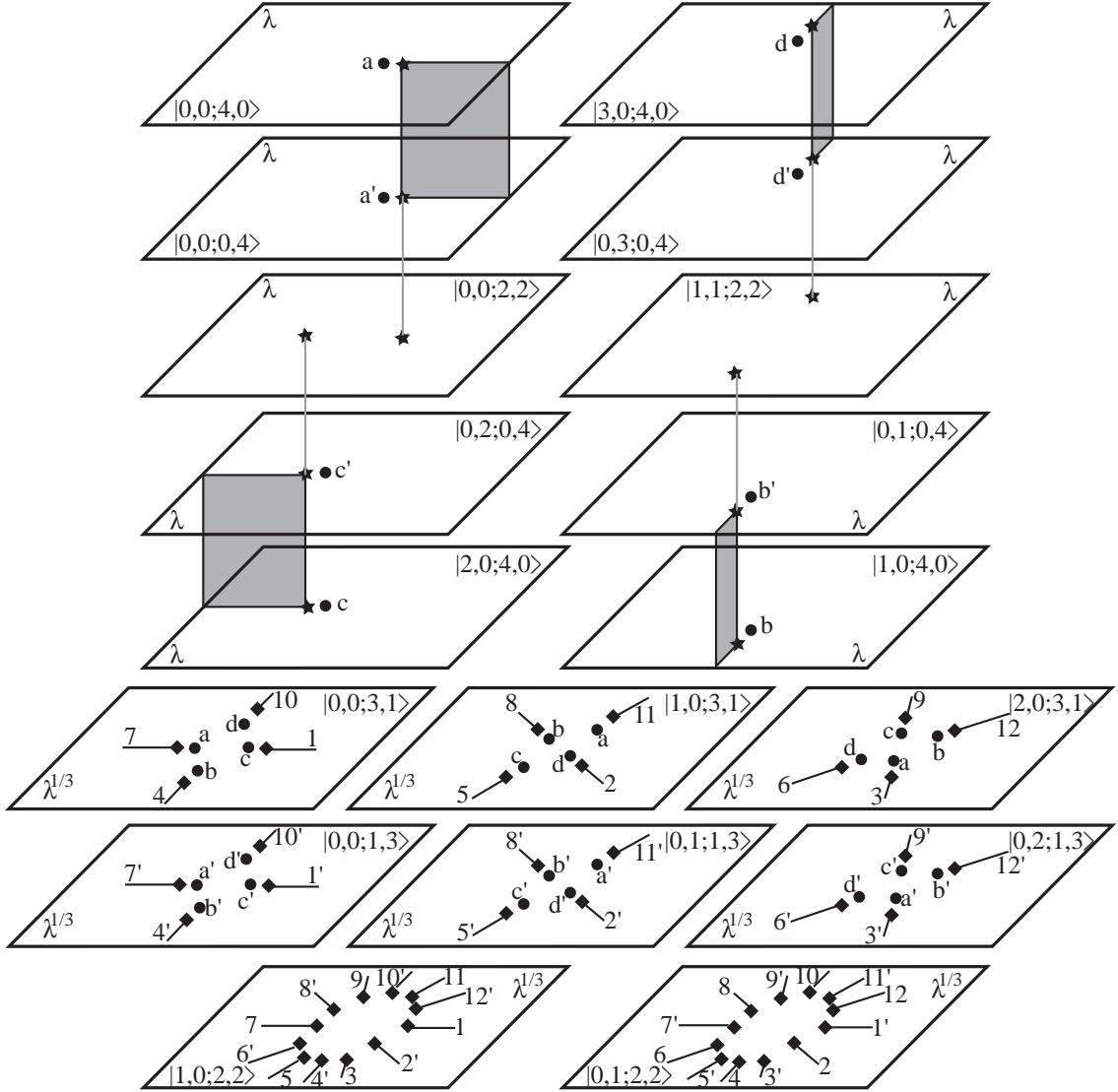,width=15cm}
}
\caption{Sketch of the quantum parameter space $\mq$ for gauge group 
$\uuuu$. The sheets are parametrized by $\la$ or $\la^{1/3}$. The
dots, squares and stars represent singularities with a massless
monopole (at $\la = (4/5)e^{-ik\pi /2}$), a massless glueball (at
$\la^{1/3}=(3^{1/4}/2^{1/6})e^{-ik\pi/6}$), or both (at $\la =
e^{-ik\pi/2}$), respectively. Due to the complexity of the diagram, we
have not been able to represent explicitly all the identifications
between sheets. It is understood that singularities and branch cuts
with the same label are identified.
\label{fig2}}
\end{figure}

We could proceed to the study of higher ranks by using the same ideas.
For example, the twenty Coulomb sheets of the ${\rm U}(5)$ theory are
separated in two sets of ten related to each other by a $2\pi$ shift
in $\theta$. The couples of integers $(N_{1},N_{2})$ in each set are
$(1,4)$ and $(4,1)$ ($2+2$ states), and $(2,3)$ and $(3,2)$ ($3+3$
states). From this we deduce the classical polynomial for $u$,
analogous to (\ref{Qcl}),
\be Q_{10,\, \rm cl}(u) = (u-1)^{2}(u-2)^{3}(u-3)^{3}(u-4)^{2}=0\, .\ee
The quantum equation can be found to be
\be Q_{10}(u) = Q_{10,\, \rm cl}(u) - {96 g^{5}\La^{5}\over m^{5}} 
(u-3)^{2} (u-2)^{2} (2u-5) - {32 g^{10}\La^{10}\over 
m^{10}}=0\, .\ee
This shows that the ten sheets are related to each other through 
branch cuts. Phase transitions with the confining vacua can also be 
found straightforwardly. They occur for
\be 89^{5}\la^{10} - 9281\times 19\times 2^{15}\,\la^{5} + 2^{30}=0 \ee
and make $\mq$ fully connected.

\subsection{An effective description}

It is interesting to discuss the effective description in terms of the
glueball superpotential $W(S_{i})$. At small $S_{i}$, the calculation
of $W$ starts by choosing a classical vacuum around which one expands
in terms of planar Feynman diagrams \cite{KKK}. As was explained by
using the matrix model in the Appendix of \cite{f1}, the equations of
motion $\d W =0$ have automatically $\prod_{j} N_{j}$
solutions. In our case, the solutions describe automatically the
vacua $|k_{1},k_{2};N_{1},N_{2}\rangle$ for given $N_{1}$ and $N_{2}$ 
and any $k_{1}$ and $k_{2}$.

A natural question is whether $W(S_{i})$ is also able to describe the
smooth interpolations between different sheets, that correspond to
different classical limits, and thus to different integers $N_{j}$.
The most basic example was studied in \cite{f1}, where it was shown
that the interpolation between the vacua $|k,0;N,0\rangle$ and
$|0,k;0,N\rangle$ (which is a strong coupling effect in the theory
with superpotential (\ref{Wtree})) is correctly described by $W(S)$.
We believe that this is a completely general
phenomenon.\footnote{This seems to
contradict a claim made in \cite{CSW}.} In the case of $\uuuu$, for
example, the interpolation between the eight Coulomb components can be
described by $W(S_{1},S_{2})$. A direct way to understand this is that
the expectation values $\langle S_{1}+S_{2}\rangle=\langle S\rangle$
relevant to those components and deduced from $\d W=0$ must satisfy the
equation (\ref{seq}) that describes the interpolation between the
different sheets. Of course, this is a genuine
non-perturbative effect that cannot be seen in the expansion of
$W(S_{i})$ in powers of $S_{i}$.
\vfill\break

\section{Conclusion}

For the first time it is possible to compute exactly and
systematically the quantum parameter spaces for a very large class
of $\nn =1$ supersymmetric gauge theories. This is an important step
forward compared to the more conventional discussions of quantum moduli
spaces, which are plagued by the presence of massless scalars. Many 
aspects remain to be uncovered. In the $\uN$ model with one adjoint 
Higgs on which we have focused, it would be desirable to better 
understand the structure and the r\^ole of multicritical points. 
There will be analogues of Argyres-Douglas points and also 
generalizations of the massless glueball points \cite{f1}. The study 
of other gauge groups and/or matter contents is likely to produce new 
interesting results. In particular, it is tantalizing to study 
chiral models.

As emphasized in \cite{f1}, very interesting phenomena are also
associated with the spectrum of domain walls and the large $N$ limit.
It was shown in particular that the large $N$ expansion can break down
near strong coupling singularities, providing new examples of a
phenomenon first discussed in \cite{F1,F2,F3,F2D,Fnp}.
The most intriguing aspect
is that it is possible to define double scaling limits, generalizing
the old approach to non-critical strings \cite{MM} to the case of four
dimensions \cite{Frevue}.

At small $S_{i}$, the Dijkgraaf-Vafa matrix model is a very simple
theory of D-branes \cite{KKK} and open/closed string duality. However,
by far the most interesting physics occurs in regimes where the
perturbative approximation to the matrix model breaks down.
At large $S_{i}$, a description in terms of closed strings
only is no longer possible. Open strings must appear \cite{F2}, and
the double scaling limits considered in \cite{F3} should correspond to
a continuum limit for these open strings. An explicit description of
this ``enhan\c con'' mechanism \cite{pol} in terms of the matrix model
would certainly provide important new insights.

\end{document}